\documentclass[journal]{IEEEtran}
\IEEEoverridecommandlockouts
\usepackage[english]{babel}
\usepackage{graphicx,multirow}
\usepackage{cite}
\usepackage{soul}
\usepackage{subfigure}
\usepackage{units}
\usepackage{siunitx}
\usepackage{booktabs}
\usepackage{epstopdf}
\usepackage{array}
\usepackage[hyphens]{url}
\usepackage{float}
\usepackage{xcolor}
\usepackage{amsmath}
\usepackage[ruled]{algorithm2e}
\usepackage{algorithmic}
\usepackage[section]{placeins}
\usepackage{newcent}
\begin{document}
%
\title{Performance Evaluation of SDN-Controlled Green Mobile Fronthaul Using a Federation of Experimental
Network}
%
%
%

\author{K. Kondepu$^*$, A. Sgambelluri$^*$, F. Cugini$^+$, P. Castoldi$^*$, R. Aparicio Morenilla$^\#$, \\D. Larrabeiti$^\#$, B. Vermeulen$^\delta$ and L. Valcarenghi$^*$

\thanks{L. Valcarenghi (email: luca.valcarenghi@santannapisa.it), K. Kondepu, A. Sgambelluri and P. Castoldi are with $^*$Scuola Superiore Sant'Anna, Pisa, Italy. F. Cugini is with $^+$CNIT, Pisa, Italy. R. Aparicio Morenilla and D. Larrabeiti are with  $^\#$Universidad Carlos III de Madrid, Madrid, Spain. B. Vermeulen is with  $^\delta$imec/Ghent University, Gent, Belgium.}
\thanks{Manuscript received \today; revised  \today}
}
\maketitle

\begin{abstract}

When evolved NodeB (eNB) flexible functional split is implemented in
Virtualized Radio Access Network (V-RAN) 5G systems, fronthaul
connectivity between the virtualized network functions (VNFs) must be
seamlessly guaranteed.

This study proposes the utilization of Software Defined Networking
(SDN) to control the mobile fronthaul.  In particular, this study
investigates the ability of the SDN-based control of reconfiguring the
fronthaul to maintain VNF connectivity when cell and optical access
turn into sleep mode (off mode) for energy efficiency purposes.

The evaluation of the proposed scheme is performed by federating two
remote experimental testbeds. Results show that, upon cell and optical
access turning on and off, the fronthaul reconfiguration time is
limited to few tens of milliseconds.

\end{abstract}


%
\IEEEpeerreviewmaketitle

\section{Introduction}
%
%
%
%

\IEEEPARstart{A}{}\label{intro}s the utilization of Internet and, as of
consequence, of communications networks is rapidly increasing much
attention is posed on communications network energy efficiency.
In parallel, Software defined networking (SDN)~\cite{Akyildiz2014} has
emerged as a strong candidate to improve the control of
telecommunication networks also in the context of mobile crosshauling
(i.e.,front/backhauling).

On the energy efficiency side, solutions for reducing the energy
consumption of macro-cells as well as small-cell have been
proposed~\cite{Qiaoyang_CoRR_2013,Boon_GC_WS2014}. Similarly
solutions for reducing the energy consumption of systems for
implementing the RAN have been proposed. For example,
energy saving solutions for Passive Optical Networks (PONs) have been targeted by international
standardization associations (i.e., ITU-T and IEEE) and research
institutions~\cite{IEEENetwork_2012}. In particular, solutions have
been devised for decreasing PON customer premises equipment (CPE)
(i.e., Optical Network Units --- ONUs) energy consumption in
10-gigabit capable PONs (e.g., XG-PON)~\cite{IEEENetwork_2012}. In
Time and Wavelength Division Multiplexed (TWDM) PONs, energy savings
also at the Optical Line Terminal (OLT) can be achieved by turning OFF some of the OLT
transceivers during low traffic periods and by sharing the fewer
active wavelengths among the ONUs equipped with tunable
transceiver~\cite{YiranOFC2012,VeenECOC2013,TaguchiECOC2013,
KanekoOFC2014}. However, all the solutions trade a decrease in
energy consumption with and increase in delay experienced by
the transported data.

On the SDN side solutions are not only proposed by academia but are
also supported by the industry~\cite{Juniper_2014,Bernardos_2014,Cvijetic_2014_Optics_Express}. For
example, in~\cite{Cvijetic_2014_Optics_Express} an SDN-controlled
optical topology-reconfigurable mobile fronthaul is proposed to carry
bidirectional coordinated multipoint (CoMP) flows between mobile
cell sites and baseband units (BBUs). The proposed solution, evaluated
in a local testbed, achieves end-to-end packet delay in the order of
few microseconds but the topology reconfiguration time is in the order
of milliseconds. In~\cite{Cvijetic_2014}, SDN and OpenFlow are
extended to control an optical access/aggregation network and
implement software-defined OLT and software defined ONUs. In the considered solution, part
of the optical spectrum unutilized by the PON is reutilized for
providing Orthogonal Frequency Division Multiple Access (OFDMA) mobile
backhaul (MBH). Other studies in the literature prove the feasibility
of an SDN-based approach for provisioning multi-technology
multi-tenant connections~\cite{Vilalta_OFC2015}. In parallel, a scheme
for flexible networking is presented in~\cite{Schrenk2015} for the
mobile front-and/or backhaul through reconfigurable nodes, where the
capacity of the mobile crosshaul can be distributed among data centers of
cloud radio access network (C-RAN) to adapt to actual load conditions
at BBUs. In addition, in~\cite{Chitimalla2017,Valcarenghi_JOCN_2017}, a
qualitative and quantitative analysis of fronthaul reconfiguration
techniques along with a study of advantages/disadvantages of
Ethernet-based fronthaul solutions is presented.

Some studies already proposed several architectures for the
utilization of C-RAN together with SDN and Network Function
Virtualization (NFV) for mobile crosshaul~\cite{Costa_EuCNC_2015}.
However their initial evaluation focused mainly on the coordination
between function migration and network reconfiguration.

This study proposes the utilization of SDN to control
mobile fronthaul when energy efficient schemes are
implemented not only at the eNB but also in part of the RAN. In
particular, this study investigates the ability of the SDN-based
control of coordinating cell and optical access device turn on/off with the
reconfiguration of the aggregation segment of the fronthaul for
providing seamless connectivity between User Equipments (UEs) and
virtualized, but static, mobile network functions.

In this paper two federated testbed facilities are utilized to
evaluate the performance of the proposed scheme. The experimental
results show that, upon cell and optical access on/off, the fronthaul
reconfiguration time is limited to few tens of milliseconds.

\section{Federation of Heterogeneous Research Network Testbeds: UltraAccess and Virtual Wall}
\label{sec:1}

Novel networking concepts are being created in research laboratories
all over the world, and demonstrated in testbeds whose utilization is
constrained to local researchers. Usually, these network testbeds
involve unique prototype components and measurement equipment. A wider
impact and analysis of the possibilities of the new technology could
be achieved if these testbeds are open to other researchers through an
advanced Internet service. Moreover, the federation of a large amount
of disperse testbed facilities can make it possible for multiple
researchers to experiment with future Internet protocols and
applications on a big computation grid for a given period of
time. This is one of the goals of Global Environment for Network
Innovations (GENI) in USA and Future Internet Research and
Experimentation (Fed4FIRE)~\cite{F4F} in Europe.  Both initiatives aim
to create a large-scale virtual laboratory for networking and
distributed systems research and education.

\subsection{The federation: Fed4FIRE}
\label{sec:2}

A number of projects aim to build a cross-national facility to enable
experimentally driven research in different parts of the world. Most
of them are focused on a single research community. Fed4FIRE (2012-2016), consolidated
in Fed4FIRE+ (2017-2021) is an EU project
that intends to build an open, accessible and
reliable framework for the federation of Internet research
infrastructure across community borders.  Sample community domains
include optical networking, wireless networking, software defined
networking, cloud computing, grid computing, and smart cities.


 All these heterogeneous communities, including next-generation optical
networking, were involved in the project to guarantee compatibility and sup-
port of heterogeneous infrastructure and experimentation requirements. The
keys of this target framework are:

\begin{enumerate}
  \item open experiment life-cycle management software;
  \item experimental measurement and monitoring tools;
  \item trust and security mechanisms, and
  \item advanced inter-testbed connectivity services.
\end{enumerate}

However, to achieve the federation of optical networking testbeds
several issues must be solved. Most optical laboratory devices do not
feature the programmability required to embed federation software, and
lack the virtualization capabilities required to enable secure access
to the infrastructure to external users. Consequently, those elements
of a testbed that do not support this functionality need to be
assisted by a computer. We have focused on this scenario and have
integrated an optical access network testbed called UltraAccess as
powerfully as possible: a Wavelength Division Multiplexed Passive
Optical Network (WDM-PON) at Carlos III University (UC3M). This
testbed is covered in the next section.

\begin{figure}[htb]
\centering
\includegraphics[width=0.9\columnwidth]{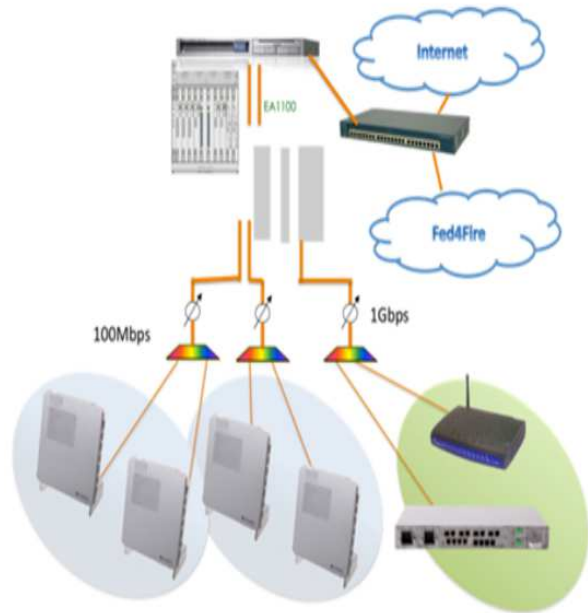}
\caption{WDM-PON testbed at U3CM}
\label{fig:pcu3cm}
\end{figure}
\subsection{The federated testbed: UltraAccess at UC3M}
UltraAccess testbed consists of a fully configurable WDM-PON network,
including a WDM-PON OLT, a multipoint fiber infrastructure equipped
with an arrayed waveguide grating (AWG), a set of Optical Network Terminals (ONTs) and high-end
Fed4Fire stations, providing end-users with a dedicated capacity of
100-1000Mb/s. The systems allows the configuration of advanced QoS
features, traffic engineering and virtual LANs. The testbed
features a seeded WDM-PON as standardized by ITU-T G.698.3. In the
seeded WDM-PON approach, a broadband light is sliced by an AWG and injected into an anti-reflection coated
Fabry-Perot laser to create ``seed" signals that transmitters can lock
onto. This technique is very attractive due to its simplicity and
cost-effectiveness, since the ONTs are colourless (i.e., they can
transmit/receive on any wavelength).

\subsection{The federated testbed: Virtual Wall at iMinds}

At the time of experimentation, iMinds Virtual wall has two kinds of set-ups: Virtual Wall 1 with 200
physical servers (e.g., 100 x quadcore, 100 x eight cores), and
Virtual wall 2 with 100 physical servers (e.g., 100 x 12 cores). All
servers were equipped with a management Ethernet interface, and
multiple (4 to 6 for the quad cores, up to 11 for the 12-cores)
Ethernet interfaces that can be used for experimentation
purposes. Both the Virtual Walls are controlled by testbed management
software \emph{jFed} based on Emulab. Multiple operating systems are
supported (e.g. Linux (Ubuntu, Centos, Fedora), FreeBSD, Windows 7) on
each node, and some of the nodes are connected to an OpenFlow switch
to do OpenFlow experiments that supports software OpenFlow switches
(OVS) and real OpenFlow switches.

\section{SDN-Controlled Energy Efficient Mobile Fronthaul Experiment}

\subsection{SDN-controlled Mobile Fronthaul Architecture}

\begin{figure}[htbp]
\centering
\includegraphics[width=1.0\columnwidth]{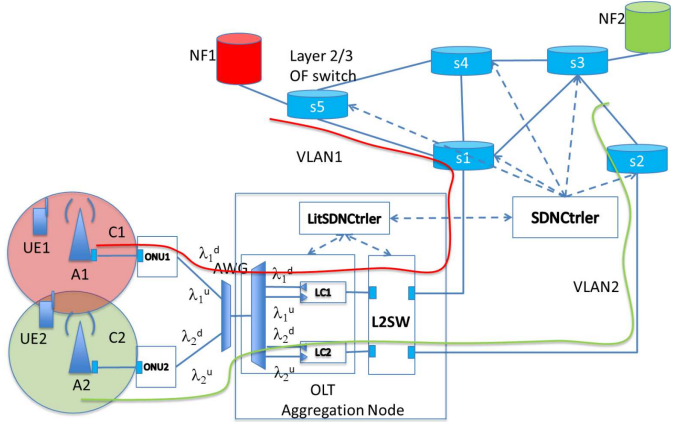}
\caption{SDN-controlled crosshaul architecture.}
\label{fig:architecture}
\end{figure}

Figure~\ref{fig:architecture} depicts the considered SDN-controlled crosshaul
architecture. It consists of a WDM-PON access and an OpenFlow based layer 2/3
aggregation network. Each antenna $A_i$ is connected to an ONU (i.e.,
ONU$_i$). Toward the aggregation network, each OLT Line Card (LC$_i$) is connected to an interface of an Ethernet
OpenFlow (OF) Layer 2 Switch (L2SW), included within the aggregation
node, that is, in turn, connected to the aggregation network. The OF
L2SW is controlled by the aggregation node controller, implemented as
a light version of an SDN controller (i.e., LitSDNCtrler) as proposed
in~\cite{Kondepu_JOCN_2017}. The aggregation network consists of
Ethernet OF Layer 2/3 switches and it is controlled by another SDN
controller (i.e., SDNCtrler). It is assumed that the chosen crosshaul
functional split allows to carry both fronthaul traffic (digital radio
signals transmitted from the Remote Radio Head (RRH) to the Base Band
Units (BBU) in the C-RAN) and backhaul traffic (regular data packets)
over a packet-switched network with QoS support. Mobile RAN functions are generally indicated as $NF_i$ and the
evolved NodeB (eNB) is assumed to support cell on/off for energy
saving purposes.

To improve the energy efficiency of the crosshaul this paper proposes the
contemporary sleep mode (i.e., on/off) of the WDM PON devices
connected to the cell implementing on/off. This implies the
contemporary reconfiguration of not only the crosshaul optical access
network but also of the crosshaul aggregation network. In particular,
every time the cell changes its status the OLT is notified and it
initiates the procedure for turning on/off the ONU and the LC to which
the cell is connected. Contemporarily, both the LitSDNCtrler and the
SDNCtrler are notified of the changes and perform the necessary
reconfigurations to allow the UE bearers to reach their original
destinations (i.e., $NF1$ and $NF2$).

\subsection{SDN-controlled Energy Efficient Mobile Fronthaul Implementation in Federated Testbeds}

\begin{figure*}[htb]
\centering
\includegraphics[width=0.85\textwidth]{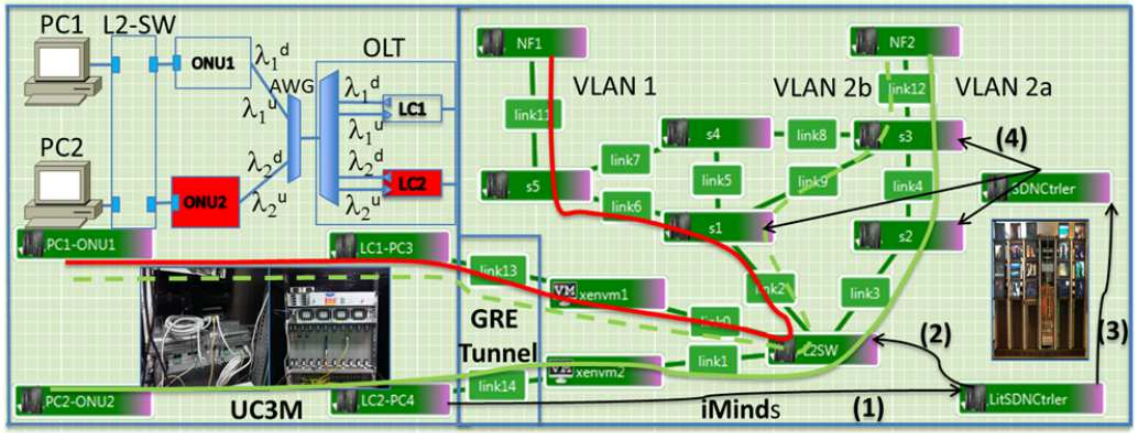}
\caption{Federated testbed setup.}
\label{fig:testbed}
\end{figure*}

In this study the architecture depicted in Figure~\ref{fig:architecture}
is implemented in two federated testbeds, depicted in
Figure~\ref{fig:testbed}, provided by the Fed4Fire project. The tesbeds
are a WDM PON testbed located at UC3M, representing the crosshaul optical
access segment, and an OpenFlow Ofelia island at iMinds, representing
the crosshaul aggregation network. The WDM PON, features two pair of
wavelengths with a capacity of 1 Gb/s each. The OpenFlow Ofelia island
consists of Ethernet Open vSwitches (OVS) $s_i$ interconnected by 1-5
Gb/s links. Two PCs on the ONUs side of the WDM PON emulate the UEs
while the layer 2 switch (i.e., L2SW) interconnecting the two PCs
with the two ONUs emulate the possibility for the UE to be connected
to either antenna. The layer 2 switch (i.e., L2SW) of the aggregation
node is an OVS as well. The LitSDNCtrler, SDNCtrler, and all the OVS
are implemented in the OpenFlow Ofelia island at iMinds. The two
testbeds are connected by means of Generic Routing Encapsulation (GRE)
tunnels through the public Internet between two PCs connected to the
OLT LCs and two Xen virtual machines (xenvm$i$) at iMinds. The tunnels
carry both data and control communications at a rate of 100Mb/s. The
signaling between the OLT and the LitSDNCtrler is performed in-band
while the signaling between the LitSDNCtrler and the SDNCtrler is
performed out-of-band through a direct connection.

\section{Utilized tools}

\subsection{Experiment Management and Control}

\begin{figure}[htbp]
\centering
\includegraphics[width=1.0\columnwidth]{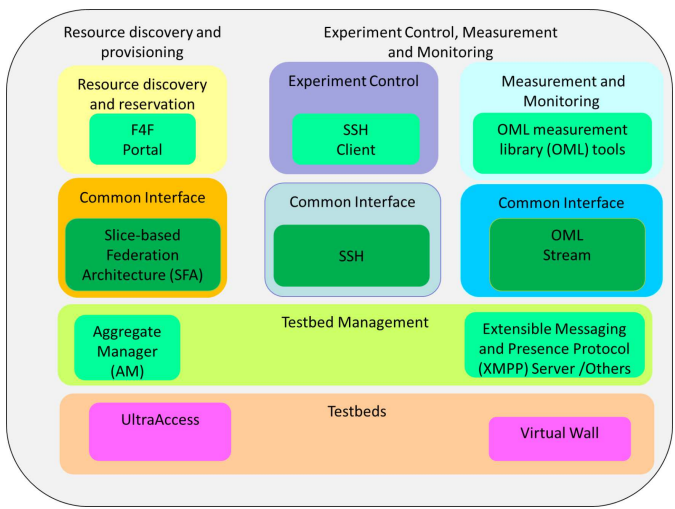}
\caption{Experiment design overview}
\label{fig:ed}
\end{figure}

Figure~\ref{fig:ed} shows the testbeds and the basic Fed4FIRE tools
used to perform the experiment. The experiment uses several resource
reservation, experiment control, and measurement and monitoring tools
including common interfaces protocols. Here, the F4F portal offers
access to the required Fed4FIRE tools and testbeds (e.g., jFed).  The
jFed tool was used to reserve the resource from both the testbeds
(i.e., UltraAccess and iMinds Ofelia island) by using slice-based
federation architecture (SFA) interface. During the experiment, the
resources of the testbeds (i.e., wavelengths, VLANs, ONUs, LCs, and
L2SWs) were controlled by experiment control (EC) with the help of
SSH protocol. The resource controller (RC) (i.e., SSH) is capable
enough to interact with all the resources in the testbeds.

The OML framework is used for measurement and monitoring the
testbeds. OML uses databases running on the aggregate manager
(AM). OMF is a framework that provides a set of tools to describe,
instrument and execute experiments and collect their results and a set
of services to manage and operate testbed resources. This framework
basically requires an experiment controller (EC) that processes an
OEDL (OMF Experiment Description Language) description of the experiment scenario and controls the required
nodes and a resource controller (RC) daemon at every resource, which
receives messages from the EC and executes the commands.

\subsection{Measurement Collection: the Orbit Management Library (OML)}

Another important aspect that is tightly related to experiment and
resource control is measurement collection. Our testbed provides this
feature through OML. OML is a generic software framework for
measurement collection. It allows the experimenter to define the so-called measurement points (MPs) inside new or pre-existing
applications so that these MPs generate measurement streams (MSs) that
can be directed to a central collection point, and stored in
measurement databases. It is composed of two entities. The first one
is an OML server responsible for collecting and storing measurements
inside an SQLite3 or PostgreSQL database. The UltraFlow Access testbed
offers a local SQLite3 database so that experimenters can direct their
measures there.

The second one is an OML client library that essentially provides a C
API to be used inside applications. There are also native
implementations in Python (OML4Py) and Ruby (OML4R) as well as third
party libraries for Java (OML4J) and Javascript/WebSocket
(OML4JS). The OML group offers both OML-enabled applications and good
tutorials for OML client application programming.

The other tools that were used to measure bandwidth and latency such as:
\emph{thrulay} traffic generator~\cite{Lie2005} tool was used to measure
the bandwidth between the source and destination node. In addition, it
also provides latency with different quantiles. \emph{tshark} is a network
protocol analyzer tool was used to measure the network reconfiguration time by
capturing packets as described in Section~\ref{eval}. \emph{Ping} tool was used
to measure the round-trip time between the source node and destination node. It was also
used to test the tunnel delay between UC3M and iMinds testbeds.

\subsection{Accessing Fed4FIRE resources: jFed}

Fed4FIRE federated resources and in particular UltraAccess testbed,
accept users from Fed4FIRE trusted identity providers.  They can
configure experiments interconnecting resources from multiple testbeds
at the same time, reserve and access them via Fed4FIRE tools such as
jFed~\cite{JFED}. In order to use this tool, users need to provide a
user certificate and a password to log in.  They can get an account
and download their certificates from iMinds authority
provider~\cite{Auth}. Once logged in, users can set up their
experiments by choosing which types of resources and from which
testbeds, configure those resources (operative system, software to be
installed, network configurations, measurement options, etc), launch
the experiments and access the resources, just by double clicking on
their icons. UltraAccess testbed offers a detailed tutorial on how to
perform all the process right from the first step, the user getting an
account, up to performing some simple experiments including
measurements~\cite{UAUC3M}. The other important feature of jFed is
\emph{jFed timeline}, it is used for executing commands instantaneously on
multiple nodes or to execute commands based on a timeline.


\section{Evaluation Scenario and Results}
\label{eval}
\subsection{Day/night cell ON/OFF}
The performed experiment is run by means of the jFed experiment
toolkit, depicted in Figure~\ref{fig:testbed}, provided by iMinds which is
used to access the federated testbeds. Initially, two VLANs (i.e.,
VLAN1 and VLAN2a) are established to carry the traffic between the
antenna which the UEs are connected to and the respective network
function locations (e.g., $UE1$ is connected to antenna $A1$ of cell
$C1$ that, in turn, is connected to the network function $NF1$
server). Then, a command is issued, through the OLT management
interface, to turn off one ONU and the respective LC with the aim of
emulating the turning off of the ONU and OLT LC upon cell turning
off. Specifically $ONU2$ and $LC2$ (i.e., $PC2-ONU2$ and $LC2-PC4$ in
Figure~\ref{fig:testbed}) are turned off. Upon issuing the ONU and LC turn
off command, the OLT notifies the LitSDNCtrler (1) that, in turn,
reconfigures the aggregation node L2SW (2) and it triggers the SDNCtrler
(3) to initiate the reconfiguration of the aggregation network switches (4).
In such a way the VLAN between UE2, now connected to A1
and C1, and the NF2 server is maintained. Specifically, as shown in
Figure~\ref{fig:testbed}, the VLAN path (labelled VLAN2b in
Figure~\ref{fig:testbed}) is ONU1-LC1-L2SW-s1-s3-NF2.

The considered performance parameter is the \emph{VLAN reconfiguration
time}, here defined as the time elapsing between the transit of the
reconfiguration triggering message sent by the OLT to the LitSDNCtrler
through the L2SW (in-band signaling) and the the detection of the
first successive \emph{ping} reply from the NF2 server at the
L2SW. Note that the additional constant delay due to the GRE tunnel is
not taken into account. In addition, the contribution of the ping
round trip time can be considered negligible because all the involved
devices (i.e., OVS, controllers, and servers) are located in the same
local network. The VLAN reconfiguration time measurement is performed
as follows: i) \emph{ping} is continuously run between xenvm1 and NF2
server with packet interval of 1ms; ii) reconfiguration request
commands sent by the OLT to the LitSDNCtrler are monitored at the OVS
implementing the L2SW at iMinds (in this way the L2SW becomes a
synchronization point for the measurement); iii) similarly,
\emph{ping} replies from the NF2 server are monitored at the L2SW. The
monitoring is performed through the \emph{tshark} tool installed in
the L2SW.

\begin{figure*}[htbp]
\centering
\includegraphics[width=1.0\columnwidth]{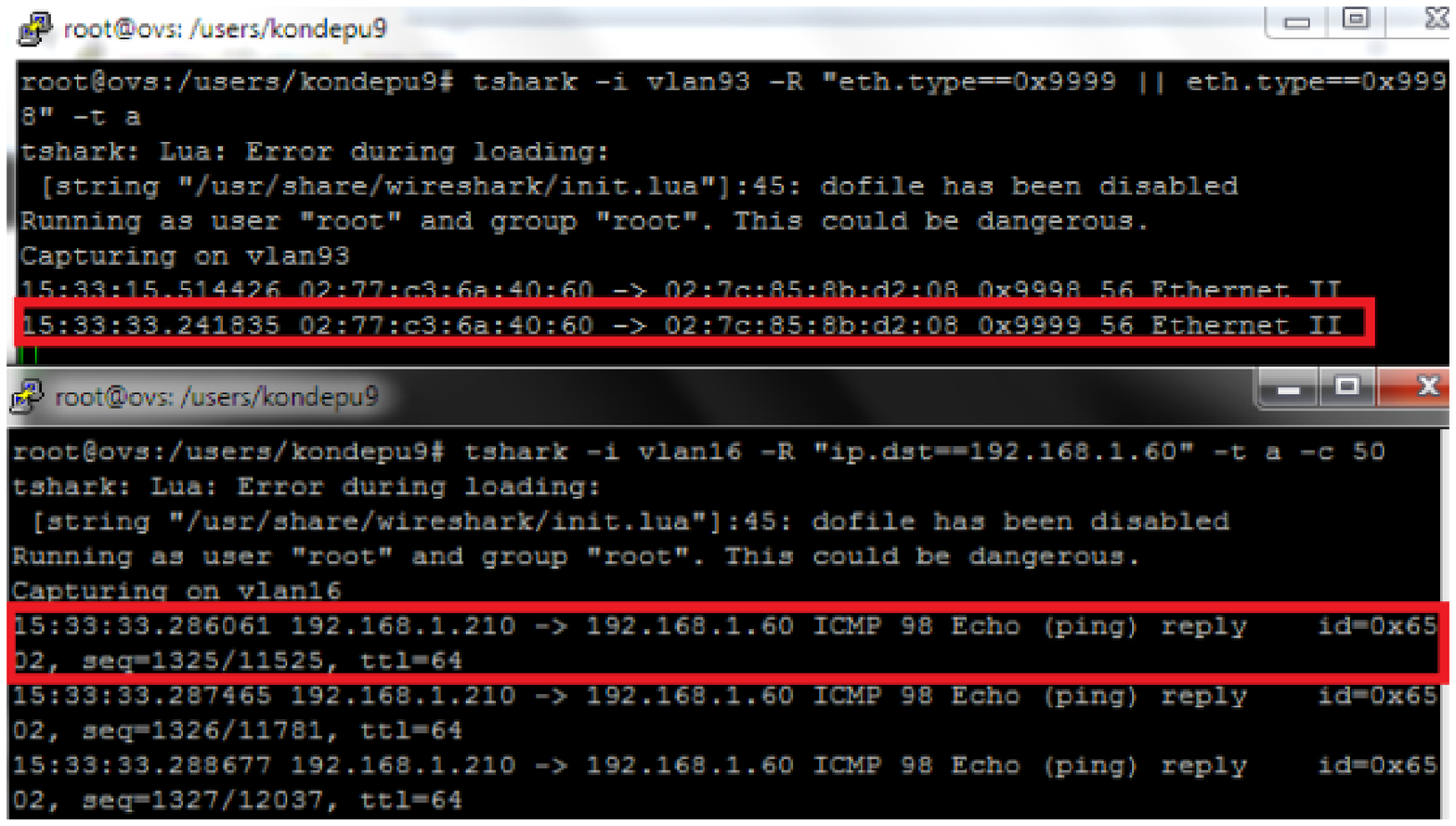}
\caption{Reconfiguration timestamp capture.}
\label{fig:capture}
\end{figure*}

Figure~\ref{fig:capture} shows the timestamp of the arrival, to the L2SW, of
the reconfiguration triggering message sent by the OLT to the
LitSDNCtrler and the timestamp of the arrival of the first successive
ping reply from NF2 server to the L2SW (red colored rectangles),
after about $45 ms$. The control plane message exchange for network
reconfiguration is the major contributor to the VLAN reconfiguration time
while the L2SW reconfiguration time contributes only for few tens of microseconds,
as reported in~\cite{ValcarenghiICTON2015}.

\begin{figure}[htbp]
\centering
\includegraphics[width=1.0\columnwidth]{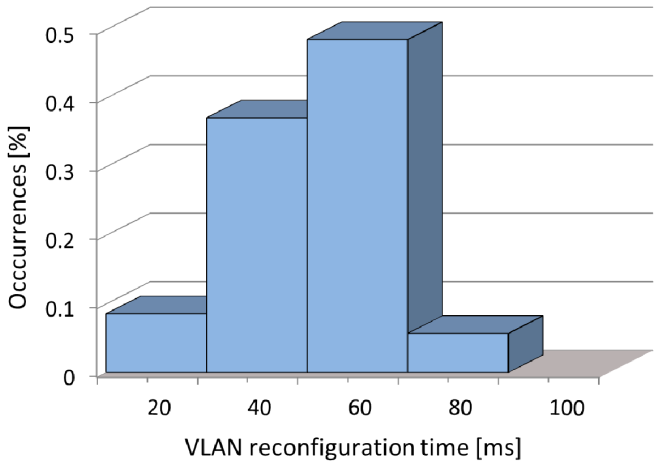}
\caption{VLAN reconfiguration time sampled pmf.}
\label{fig:en}
\end{figure}

Figure~\ref{fig:en} shows the sampled probability mass function
of the VLAN reconfiguration time during a single
experiment with duration of 3600s (60 reconfigurations). In
particular, around 50\% of network reconfigurations take between 40ms
and 60ms. Thus, it is experimentally proven that a VLAN
reconfiguration time of few tens of milliseconds is achieved by the
proposed SDN-controlled mobile crosshaul. In addition, crosshaul energy
savings proportional to the cell off time and the difference between
the energy consumed by the ONU and the OLT during on and off periods
are possible.

In the current experiment energy efficiency was not evaluated but an estimate
can be provided by considering how often the cell is turned ON/OFF. Without
considering the cell energy consumption, and by assuming that a cell is OFF
for one fourth of a day (e.g., during the night), if the power consumed by
an OLT LC when it is working is 6W and when it is OFF is 4.2W (our assumption)
and the power consumption of the ONU when it is ON is 3.2W and when it is OFF
is 2.3W~\cite{ccecbe2013}, the energy savings that can be potentially achieved
are about 4\%. In general, they can be as computed as follows:

The power consumed when both LCs of OLT and both ONUs are always ON is:
\begin{equation}
P_{ON}= 2(P^{OLT}_{ON} + P^{ONU}_{ON})
\label{e:pon}
\end{equation}

The power consumed when one LC-ONU pair is ON, and the other LC-ONU pair is OFF (as depicted in Fig.~\ref{fig:testbed}) is:
\begin{equation}
P_{OFF} = P^{OLT}_{ON} + P^{ONU}_{ON}+P^{OLT}_{OFF} + P^{ONU}_{OFF}
\label{e:poff}
\end{equation}

>From Eq.~(\ref{e:pon}), and Eq.~(\ref{e:poff}), the average energy savings for the considered scenario can be computed as:
\begin{equation}
\eta= 1- \frac{T_{ON}P_{ON} + T_{OFF}P_{OFF}}{P_{ON}(T_{ON} + T_{OFF})}
\label{e:Saving_equation_bound}
\end{equation}

\subsection{Fast cell ON/OFF}

\begin{figure}[htb]
\centering
\includegraphics[width=1.0\columnwidth]{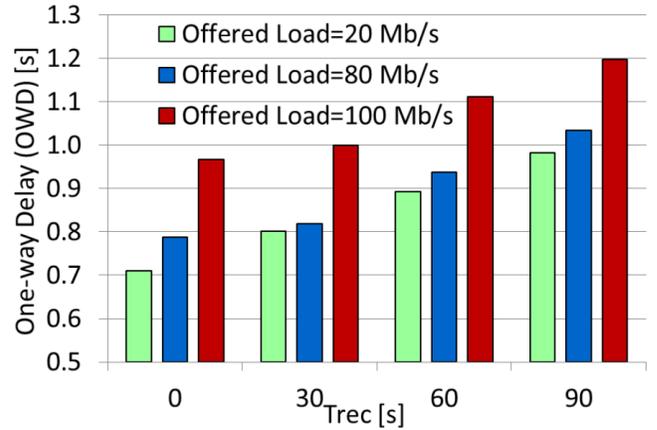}
\caption{Average frame delay as a function of the reconfiguration period Trec.}
\label{fig:trec}
\end{figure}
In this section the performance when a fast cell ON/OFF scheme is implemented are reported
in Figure~\ref{fig:trec}. Thrulay traffic generator~\cite{Lie2005} tool is used to measure
the end-to-end average frame delay (i.e., UC3M PC2 to iMinds server2) while periodically
turning ON/OFF LC2. When LC2 is turned OFF the traffic is aggregated to LC1, as shown in
Figure~\ref{fig:testbed}. Thrulay server modules are running at both the servers of iMinds,
and Thrulay client modules are running at PCs that are attached to both the ONUs. UDP traffic
is generated from the clients with different traffic loads, packet size is set to $1400B$, and
UDP buffer size is set to $262142B$. The experiment duration is $2400s$ for all the considered
scenarios. The automated reconfiguration request commands are periodically sent from the PC4 attached
to LC2 (i.e., LC2-PC4as shown in Figure~\ref{fig:testbed}), to the LitSDNCtrler and they are monitored
at the OVS implementing the L2SW at iMinds.

The considered performance parameter are \emph{One-way Delay} and \emph{One-way Packet Loss Ratio}.
The \emph{One-way Delay} is defined as the time it takes for a packet to reach its destination.
The \emph{One-way Packet Loss Ratio} is defined as ratio between the number of received packets at
the destination (e.g, NF2) and the number of packets sent at the source (e.g., PC2).

Figure~\ref{fig:trec} shows the one-way delay as a function of reconfiguration period $T_{rec}$. As expected,
the one-way delay increases as Trec increases, because the number of packets buffered during the reconfiguration
period increases. Table~\ref{t:plr} shows the one-way packet loss ratio as a function of reconfiguration period $T_{rec}$.

\begin{table}
\caption{One-way Packet Loss Ratio [\%]}
\label{t:plr}
\setlength{\tabcolsep}{2.5em}
\renewcommand{\arraystretch}{1.5}
\centering
\begin{tabular}{|c|c|c|c|}
\hline
Trec&\multicolumn{3}{c|}{Offered load [Mb/s]}\\ \cline{2-4}
[s] &20&80&100 \\ \hline
0&	0&	0.651&	3.907 \\ \hline
30&	0&	0.718&	4.01 \\ \hline
60&	0&	0.746&	4.018 \\ \hline
90&	0.001&	1.034&	4.198 \\ \hline
\end{tabular}
\end{table}
\vspace{-3mm}

\section{Conclusions}
This study proposed an SDN-based solution to coordinate cell on/off
with part of the mobile crosshaul on/off (and subsequent reconfiguration)
for improving crosshaul energy efficiency. Results show that, once part
of the crosshaul is turned off, communication between the user equipment
and the server hosting a specific network function is recovered after
few tens of milliseconds. Thus, the proposed solution is compatible
with a centralized coordinated Radio Resource Control (RRC) implementation
whose latency requirements are in the order of seconds.

The experiments carried out to assess the approach relied on third-party-provided
network equipment over a testbed federation platform. This shows that complex set-ups
for 5G crosshaul experimentation can be realized and used for valid experimentation,
thanks to the advances in virtualisation and testbed federation currently available to
the research community.

\section*{Acknowledgment}
This work was supported in part by the Fed4FIRE project (``Federation
for FIRE''), an integrated project funded by the European Commission
through the 7th ICT Framework Programme (318389), partly by the EU-funded
Wishful project (645274), and partly by the EU-funded 5G-Crosshaul project (671598).

\bibliographystyle{IEEEtran}
\bibliography{Fed4Fire2017}

\begin{IEEEbiographynophoto}{Koteswararao Kondepu} is currently working as a Research Fellow at Scuola Superiore Sant' Anna, Pisa, Italy. He obtained his Ph.D. degree in Computer Science and Engineering from Institute for Advanced Studies Lucca (IMT), Italy in July 2012. His research interests are 5G, optical networks design, energy efficient schemes in communication networks and sparse senor networks.
\end{IEEEbiographynophoto}
\vspace{-1cm}
\begin{IEEEbiographynophoto}{Andrea Sgambelluri} received Ph.D. degree from Scuola Superiore Sant'Anna, Pisa, in December 2015. In 2016, he got the postdoc researcher position at KTH Royal Institute of Technology (Optical Networks Laboratory (ONLab)) Sweden. Since December 2016, he is postdoc researcher at the TeCIP Institute of Scuola Superiore Sant'Anna, Pisa, Italy. His main research interests are in the field of control plane techniques for both packet and optical networks, including SDN protocol extensions and Network Function Virtualization (NFV). He is co-author of more than 30 publications in international journals and conference proceedings, and he filed 1 international patent.
\end{IEEEbiographynophoto}
\vspace{-1cm}
\begin{IEEEbiographynophoto}{Filippo Cugini} is Head of Research Area at CNIT, Pisa, Italy. His main research interests include theoretical and experimental studies in the field of packet and optical communications, including SDN control plane, segment routing, and cloud networking. He is co-author of 15 patents and more than 200 international publications.
\end{IEEEbiographynophoto}
\vspace{-1cm}
\begin{IEEEbiographynophoto}{Piero Castoldi} (PhD in Information Technology) has been Professor at Scuola Superiore Sant'Anna, Pisa, Italy since 2001. He spent abroad at Princeton University (USA) overall about two years in 1996, 1997, 1999, 2000 and he has visited the University of Texas at Dallas, USA for two months in 2002.
He has also served as Project Manager of many initiatives of the Inter-universitary National Consortium for Telecommunications (CNIT). He is currently Leader of the "Networks and Services" research area at the TeCIP Institute of Scuola Superiore Sant'Anna. His research interests cover telecommunications networks and system both wired and wireless, and more recently reliability, switching paradigms and control of optical networks, including application-network cooperation mechanisms, in particular for cloud networking. He is an IEEE Senior Member and he is author of more than 400 publications in international journals and conference proceedings.
\end{IEEEbiographynophoto}
\vspace{-1cm}
\begin{IEEEbiographynophoto}{Raquel Aparicio} is currently working at TOUCHVIE, Madrid, Spain. She worked in Fed4FIRE EU-funded project between 2014 and 2016. Her research interests includes optical transparent networks, multimedia networks, and wireless sensor networks.
\end{IEEEbiographynophoto}
\vspace{-1cm}
\begin{IEEEbiographynophoto}{David Larrabeiti} is a professor of switching and networking architectures at U. Carlos III of Madrid since 1998. He has participated in EU-funded research projects related to next-generation networks and protocols through FP6 and FP7, like GCAP, OPIUM, Fed4FIRE or the BONE European network of excellence on optical networking. He is currently involved in the H2020 5G-Crosshaul and BlueSpace research projects, participating in the development of technology and testbeds for backhaul network design.
\end{IEEEbiographynophoto}
\vspace{-1cm}
\begin{IEEEbiographynophoto}{Brecht Vermeulen} received the M.Sc. degree in Electro-technical engineering and the Ph.D degree from the Ghent University, Belgium in 1999 and 2004, respectively. Since the start of the research institute iMinds (now merged with imec) in 2004, he leads the Technical Test centre iLab.t in Gent, Belgium where he started a.o. the local deployment of Emulab based testbeds. He leads research teams on server and network performance and on testbed federation architecture and tools, and coordinates technically the iMinds'/imec testbeds on Future Internet. In the Fed4FIRE project, he is leading the architecture work package, and is deeply involved in international federation.
\end{IEEEbiographynophoto}
\vspace{-1cm}
\begin{IEEEbiographynophoto}{Luca Valcarenghi} has been Associate Professor at the Scuola Superiore Sant'Anna of Pisa, Italy, since 2014. He published more than 100 papers in International Journals and Conference Proceedings and actively participated in the TPC of several IEEE conferences, such as Globecom and ICC. Dr. Valcarenghi received a Fulbright Reaserch Scholar Fellowship in 2009 and a JSPS "Invitation Fellowship Program for Research in Japan (Long Term)" in 2013. His main research interests are Optical Networks design, analysis, and optimization; Artificial Intelligence optimization techniques; Communication Networks reliability; fixed and mobile network integration; fixed network backhauling for mobile networks; energy efficiency in communications networks.
\end{IEEEbiographynophoto}

\end{document}